# Re-entrant Structural Phase Transition in a Frustrated Kagome Magnet, Rb$_2$SnCu$_3$F$_{12}$

Lewis J. Downie[a], Stephen P. Thompson[b], Chiu C. Tang[b], Simon Parsons[c] and Philip Lightfoot*[a]

**Rb$_2$SnCu$_3$F$_{12}$ has been studied using synchrotron powder X-ray diffraction, powder neutron diffraction and single crystal X-ray diffraction at a range of temperatures (100 – 500 K). A broad but clear phase transition, possibly of re-entrant character, is found to occur in the powder form, whereas the corresponding transition is not seen in the single crystal.**

Frustrated magnetism has been of great interest to both physicists and chemists for a long time due to the unusual effects and phenomena that it can encourage[1,2]. These can include spin glasses, spin ices [3,4] and spin liquids[5]. In the spin liquid (or 'resonating valence bond') state[6] there is such a high level of frustration that any tendency towards long-range magnetic order is suppressed, even down to 0 K. This is of great fundamental interest as such a state may help the understanding of, for example, high-temperature superconductivity.

Magnetic frustration is often manifested in crystal structures containing triangular motifs of magnetic species, such as those based on the kagome lattice. In the specific case of kagome lattices based on $S = ½$ ions (where the competition between long-range ordering and quantum fluctuation effects is strongest) there have been several recently reported examples of spin liquid-like behaviour. Key examples are the Cu$^{2+}$-based systems Herbertsmithite[7] and Kapellasite[8], and the V$^{4+}$ system [NH$_4$]$_2$[C$_7$H$_{14}$N][V$_7$O$_6$F$_{18}$][9,10]. Each of these examples has been shown to display persistent spin fluctuations down to below 50 mK. A related phenomenon, the so-called 'valence-bond solid' or VBS state, in which localised, static spin-pairing exists, but still without long-range cooperative interactions, occurs in the compound Rb$_2$SnCu$_3$F$_{12}$. This is an attractive system for exploring crystal-chemical influences on frustration due to the chemical flexibility inherent within the wider A$_2$MCu$_3$F$_{12}$ family[11]. Some members of this family are reported to have a structural phase transition below room temperature which distorts their perfect kagome lattice[12,13] and therefore affects their preferred magnetic ground state. The crystal structure of Rb$_2$SnCu$_3$F$_{12}$ has only been examined in detail at room temperature[14], although some evidence for a phase transition near 215 K has been suggested[15]. Even at room temperature, Rb$_2$SnCu$_3$F$_{12}$ does not possess a perfect Cu$^{2+}$ kagome lattice. A structural frustration, perhaps caused by the incorporation of the smaller Rb$^+$ rather than Cs$^+$ cation, leads to a lowering of symmetry (specifically a doubling of the a and b parameters of the rhombohedral unit cell) and a disorder of the fluoride sublattice, when compared to the previously well characterised analogues Cs$_2$ZrCu$_3$F$_{12}$ and Cs$_2$SnCu$_3$F$_{12}$. These models are herein referred to as 'Rhomb-1' and 'Rhomb-2' as shown in Figure 1. The lowering of symmetry is suggested to be intrinsically related to the adoption of the unique VBS ground state, since it gives rise to four, rather than one, independent Cu---Cu superexchange pathways[15]. This contrasts with the long-range antiferromagnetic order seen in the other family members.

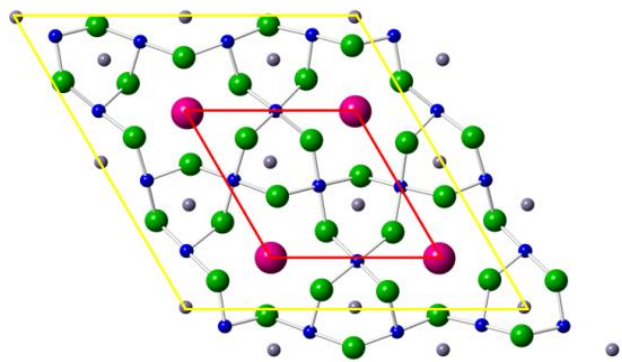

Fig 1.: Unit cell of A$_2$MCu$_3$F$_{12}$ looking down the c-axis. Red outline is the aristotype unit cell (Rhomb-1) common to Cs$_2$ZrCu$_3$F$_{12}$ and Cs$_2$SnCu$_3$F$_{12}$; yellow outline is the doubled cell (Rhomb-2) found for Rb$_2$SnCu$_3$F$_{12}$.

Considering these differences it is of interest to probe the structural behaviour of Rb$_2$SnCu$_3$F$_{12}$ at low temperature, in order to compare and contrast to the structural and magnetic behaviour of Cs$_2$ZrCu$_3$F$_{12}$.

The target compound was synthesised by an adaptation of the previously reported procedure[14]. RbF (Sigma Aldrich 99.8 %) and SnF$_4$ (Sigma Aldrich) were dried at 120 °C under vacuum for approximately 20 hours. They were then mixed and ground with CuF$_2$ (Sigma Aldrich 98%) under argon in the molar stoichiometric ratio 2:1:3 before sealing in a gold tube. The tube was then heated under flowing argon at 10 °C min$^{-1}$ to 600 °C for twelve hours before cooling at the same rate. In the case of single crystals the molar ratio was adjusted to 3:2:3 (in order to provide a flux for crystal growth), the temperature to increased 800 °C and the cooling rate slowed to ~ 3 °C hr$^{-1}$. In all cases it was very difficult to synthesise the pure Rb$_2$SnCu$_3$F$_{12}$ phase, with significant impurities being Rb$_2$SnF$_6$ and CuO. In the powder diffraction experiments we were able to adequately account for these using multiphase Rietveld refinements[16,17] – see ESI for further details.

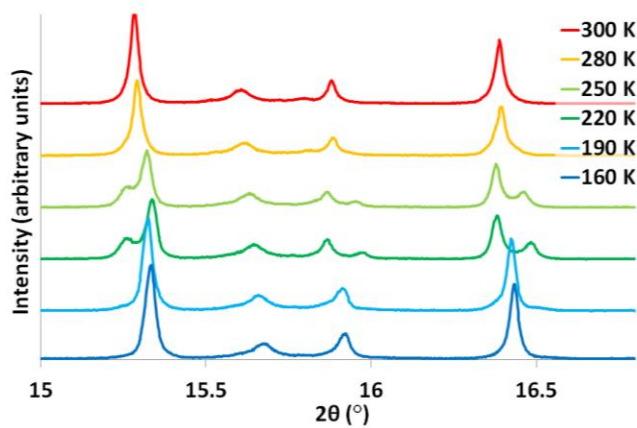

Fig 2.: Portion of synchrotron powder diffraction pattern for Rb$_2$SnCu$_3$F$_{12}$ at various temperatures.

Synchrotron X-ray powder diffraction experiments were performed at beamline I11, Diamond Light Source[18]. In this first experiment (Run 1), data were recorded at room temperature before quickly cooling to 100 K and allowing equilibration, before increasing the temperature. The room temperature data were well fitted by the proposed Rhomb-2 model (large cell, Figure 1; space group R-3, with a ~ 13.9 Å, c ~ 20.3 Å) previously derived from single crystal data[14]. Specifically it was confirmed that the doubling of the aristotype unit cell was required. At around 250 K – on cooling – however, it was found that visible splittings occurred in some of the diffraction peaks. There are two possible causes of these splittings: a phase *separation* to two very similar phases, or a phase *transition* to a single, lower symmetry phase. A thermally-induced phase separation has recently been reported in the multiferroic fluoride $K_xFeF_3$[19]. In the present case, the simplest model was tested: ie two Rhomb-2 like phases with differing lattice parameters and phase fractions. This model failed to account for the observed splittings. Initial modelling using a single lower symmetry phase was attempted using a monoclinic model which had previously been derived for the $Cs_2ZrCu_3F_{12}$ system[12]; however, this was unable to explain some superstructure peaks necessitated by the 'doubled' cell. By considering the possible distortions of the parent phase (using the ISODISTORT program[20]) a triclinic model was found to account for all the observed peaks (see ESI for full details). This model (a maximal non-isomorphic subgroup of Rhomb-2) provides the most plausible explanation of the observed data. However, due to its complexity, and the issues of sample purity, it was not possible to refine full details of this model; nevertheless, high-quality fits were obtained by refining only the lattice and profile parameters. Further cooling – to approximately 190 K – leads to the disappearance of this splitting and a return to a pattern which is well fitted by the Rhomb-2 model (Fig. 2 and 4). In order to probe this phenomenon further, and to confirm that this apparent re-entrant phase transition is not an artefact due to sluggish kinetics, the experiment was repeated, this time using both heating and slow-cooling cycles (Run 2). The sample was cooled slowly from 300 K to 165 K (below the suggested transition) and heated to 300 K again, with several data collections carried out during each ramp. In this case it was found that – visually at least – there was a clear transition from the Rhomb-2 to the triclinic phase, and *back* to the Rhomb-2 phase on both heating and cooling. Rietveld refinement showed a hysteresis in the phase transition temperature and a mixed-phase zone in both directions, both suggestive that the phase transition is 1$^{st}$ order in nature.

To obtain a definitive structure for the intermediate triclinic phase and to investigate the relationship between the low temperature and room temperature rhombohedral phases, single crystal X-ray diffraction was performed. Data were collected on two separate crystals. In the first experiment a crystal was mounted at 220 K. After data collection, the crystal was warmed to 300 K, cooled to 230 K and then warmed to 245 K. To our surprise, all three datasets, collected at 220, 230 and 245 K were indexable and refined well in the Rhomb-2 model, with R-factors of around 5% in each case (see ESI for details). It was also found that the previously reported structural disorder was present at all three temperatures. In a second single crystal experiment data were collected at 300, 245 and 140 K on cooling: in this case the quality of the refinement using the same Rhomb-2 model was inferior at 140 K to that at 245 K (R1 values of 5.62% and 3.49%, respectively). Significantly, there was no evidence for loss of three-fold symmetry, however, re-integrating the 140 K data in a much larger and lower symmetry unit cell (model 'Trig-1', space group P-3, a ~ 27.8 Å, c ~ 20.3 Å), as tentatively suggested by Matan et al.[15], did reveal some evidence for additional very weak scattering in the enlarged cell. The weakness of these possible superlattice reflections precluded any reasonable refinement. We re-iterate that this enlarged, but high-symmetry, unit cell *does not* account for the peak splittings seen in the intermediate (triclinic) phase in our powder experiments.

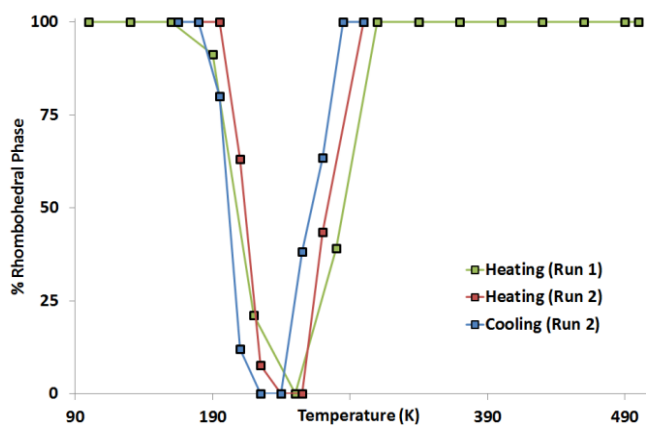

Fig 3.: Percentage of Rhomb-2 phase present in two samples.

In order to further explore the reproducibility of the powder experiments an analogous experiment was carried out using powder neutron diffraction (see ESI for details). It was found that further refinement for full structural details was prohibited due to impurity levels but at room temperature and 150 K the pattern was successfully fitted in the Rhomb-2 model and at 220 K in the triclinic cell. This confirmed the essential behaviour proposed in the synchrotron powder experiments.

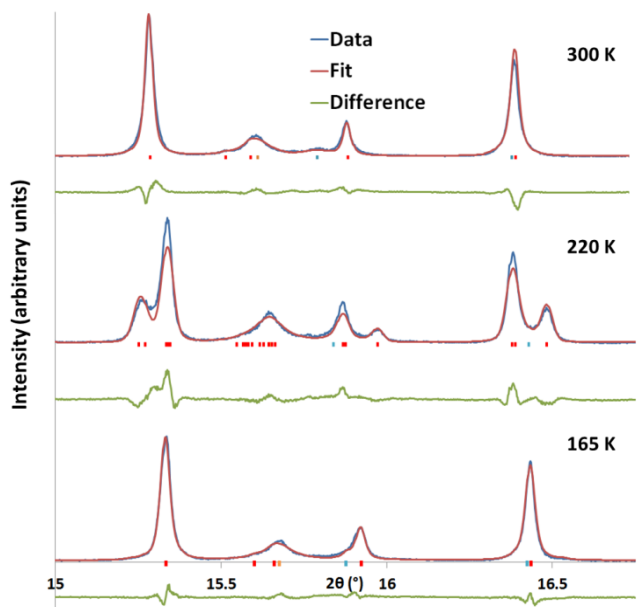

Fig. 4: Rietveld fits to the synchrotron powder data at different temperatures. In this case only single $Rb_2SnCu_3F_{12}$ phases were used (Rhomb-2 model at 300 and 165 K, triclinic model (P-1) at 220 K. Red tick-marks correspond to $Rb_2SnCu_3F_{12}$, orange to $Rb_2SnF_{12}$ and blue to $RbCuF_3$.

From the present study it is clear that powder samples of $Rb_2SnCu_3F_{12}$ show curious and hitherto unreported phase transition behaviour. The reproducible transition from rhombohedral to triclinic to rhombohedral phases on cooling and re-heating strongly suggests a re-entrant structural phase transition. Unfortunately, the exact details of each crystal structure cannot be ascertained from the powder diffraction data alone, and both the present and previous single crystal studies clearly show behaviour different to all the present powder-based studies. In particular it is not clear whether the two 'Rhomb-2' phases surrounding the triclinic phase are truly identical or whether they differ, for example, in the degree or nature of the fluoride disordering.

Purely structural re-entrant phase transitions in crystalline materials are very rare. A recent study suggests only three known examples[21]; this has now been supplemented by one further possible case[22]. Interestingly, in the latter case, the transformation has the same group-subgroup relationship (R-3 to P-1) as presented here, though it is suggested that the transition is due to competition between a weak Jahn-Teller effect and magnetic ordering; in our case the cause is more likely to be purely structural and related to the fluorine disorder.

## Conclusions

In conclusion we have found that the frustrated kagome magnet $Rb_2SnCu_3F_{12}$ shows an unusual, apparently re-entrant, structural phase transition below ambient temperature. This transition occurs reproducibly in powder form, but not in single crystals, suggesting that it is sluggish and particle/domain size dependent. An alternative, but we suggest less likely, explanation might be differences in stoichiometry between powder and single crystal samples: we have no evidence for this, but cannot rule it out due to the impurity problems in the powder sample. Although the full crystal structure of $Rb_2SnCu_3F_{12}$ has not been reported at sub-ambient temperature, the previous study of Matan[15] suggests single crystal diffraction evidence of a structural phase transition near 215 K. Moreover, a solid state $^{63,65}$Cu NMR study[23] suggests the presence of four distinct Cu sites at 5.2 K, which the authors tentatively suggest is related to the transition near 215 K. The present study does not rule out such a phase transition, in single crystal form only, but the behaviour of the powder form is certainly different. In the light of this work, further characterisation and comparison of the magnetic behaviour of single crystal versus powder samples is prompted.

## Acknowledgements

We thank EPSRC for a studentship (to LJD), Dr Aziz Daoud-Aladine for assistance with the PND measurements and The Diamond Light Source for support.

## Notes and references

[a] EaStCHEM and School of Chemistry, University of St Andrews, St Andrews Fife, KY16 9ST, UK. Fax: +44 1334 463808; Tel: +44 1334 463800; E-mail: pl@st-andrews.ac.uk
[b] Diamond Light Source, Ltd., Harwell Science and Innovation Campus, Didcot, Oxfordshire, OX11 0DE, UK.
[c] EaStCHEM and School of Chemistry, University of Edinburgh, Joseph Black Building, West Mains Road, Edinburgh, Scotland EH9 3JJ.

1. A. P. Ramirez, *Annu. Rev. Mater. Sci.*, 1994, **24**, 453.
2. A. Harrison, *J. Phys.: Condens. Matter*, 2004, **16**, S553.
3. K. Binder, A. P. Young, *Rev. Mod. Phys.*, 1986, **58**, 801.
4. S. T. Bramwell, M. J. P. Gingras, *Science*, 2001, **294**, 1495.
5. L. Balents, *Nature*, 2010, **464**, 199.
6. P. W. Anderson., *Mat. Res. Bull.*, 1973, **8**, 153.
7. P. Mendels, F. Bert, M. A. de Vries, A. Olariu, A. Harrison, F. Duc, J. C. Trombe, J. S. Lord, A. Amato, C. Baines, *Phys. Rev. B*, 2007, **98**, 077204.
8. B. Fåk, E. Kermarrec, L. Messio, B. Bernu, C. Lhuillier, F. Bert, P. Mendels, B. Koteswararao, F. Bouquet, J. Ollivier, A. D. Hillier, A. Amato, R. H. Colman, A. S. Wills, *Phys. Rev. Lett.*, 2012, **109**, 037208.
9. F. H. Aidoudi, D. W. Aldous, R. J. Goff, A. M. Z. Slawin, J. P. Attfield, R. E. Morris, P. Lightfoot, *Nat. Chem.*, 2011, **3**, 801.
10. L. Clark, J.C Orain, F. Bert, M. A. De Vries, F.H. Aidoudi, R. E. Morris, P. Lightfoot, J. S. Lord, M. T. F. Telling, P. Bonville, J.P. Attfield, P. Mendels, A. Harrison, *Phys. Rev. Lett.*, 2013, **110**, 207208.
11. See for example: T. Amemitya, M. Yano, K. Morita, I. Umegaki, T. Ono, H. Tanaka, K. Fujii, H. Uekusa, *Phys. Rev. B.*, 2009, **80**, 100406; U. Englich, C. Frommen, W. Massa, *J. Alloys Compd.*, 1997, **246**, 155; A. Le Bail, Y. Gao, J. L. Fourquet, C. Jacoboni, *Mat. Res. Bull.*, 1990, **25**, 831.
12. S. A. Reisinger, C. C. Tang, S. P. Thompson, F. D. Morrison, P. Lightfoot, *Chem. Mater.*, 2011, **23**, 4234.
13. Y. Yamabe, T. Ono, T. Suto, H. Tanaka, *J. Phys.: Condens. Matter*, 2007, **19**, 145253.
14. K. Morita, M. Yano, T. Ono, H. Tanaka, K.Fujii, H. Uekusa, Y. Narumi, K. Kindo, *J. Phys. Soc. Jpn.*, 2008, **77**, 043707.
15. K. Matan, T. Ono, Y. Fukumoto, T. J. Sato, J. Yamaura, M. Yano, K. Morita, H, Tanaka, *Nat. Phys.*, 2010, **6**, 865.
16. A. C. Larson, R. B. Von Dreele, *Los Alamos National Laboratory Report*, 1994, LAUR 86-748
17. B. H. Toby, *J. Appl. Cryst.*, 2001, **34**, 210
18. S. P. Thompson, J. E. Parker, J. Potter, T. P. Hill, A. Birt, T. M. Cobb, F. Yuan, C. C. Tang, *Rev. Sci. Instrum.*, 2009, **80**, 075107
19. S. A. Reisinger, M. Leblanc, A.-M. Mercier, C. C. Tang, J. E. Parker, F. D. Morrison and P. Lightfoot, *Chem. Mater.*, 2011, **23**, 5440.
20. B. J. Campbell, H. T. Stokes, D. E. Tanner, D. M. Hatch, *J. Appl. Cryst.*, 2006, **39**, 607.
21. M. T. Dove, *J. Phys.: Condens. Matter.*, 2011, **23**, 225402.
22. H. Kabbour, R. David, A. Pautrat, H. Koo, M. Whangbo, G. André, O. Mentré, 2012, **51**, 11745.
23. H. Tashiro, M. Nishiyama, A. Oyamada, T. Itou, S. Maegawa, M. Yano, T. Ono, H. Tanaka, *J. Phys.: Conf. Ser.*, 2011, **320**, 012052.